\documentclass[final,times]{elsarticle}

\usepackage{lineno}
\usepackage[hidelinks]{hyperref}

\usepackage{graphicx}
\usepackage{amsmath}   

\begin{document}

\begin{frontmatter}

\title{The European Strategy and Detector R\&D Program}

\author[1]{Thomas Bergauer\corref{cor1}%
  \fnref{fn1}}
\ead{Thomas.Bergauer@oeaw.ac.at}

 \cortext[cor1]{Corresponding author}
 \fntext[fn1]{The author is currently the chair of the Detector R\&D Committee (DRDC).}

\affiliation[1]{organization={HEPHY Vienna}, 
                 addressline={Nikolsdorfer Gasse 18},
                 postcode={1050}, 
                 city={Wien, Vienna}, 
                 country={Austria}}

\begin{abstract}
The latest update of the European Strategy for Particle Physics stimulated the preparation of the European Detector Roadmap document in 2021 by the European Committee for Future Accelerators ECFA. This roadmap, defined during a bottom-up process by the community, outlines nine technology domains for HEP instrumentation and pinpoints urgent R\&D topics, known as Detector R\&D Themes (DRDTs). Task forces were set for each domain, leading to Detector R\&D Collaborations (DRDs), now hosted at CERN. After an intensive period over the last months, seven DRD collaborations have been established which are now starting to set up their collaboration structures and begin to work. One is still in the preparation phase. 

In this publication, I will give an overview of the set-up process and the current status of all DRD collaborations covering detector developments in the field of gaseous detectors, noble liquid detectors for rare event searches, semiconductor detectors, photodetectors and concepts for particle ID, quantum sensors, calorimetry, electronics for HEP instrumentation and mechanical and integration aspects.
\end{abstract}

\begin{keyword}
ECFA Detector roadmap; DRD collaborations
\end{keyword}

\end{frontmatter}


\section{Introduction}
Instrumentation must not be the limiting factor for meeting the needs of the long-term European particle physics program, as outlined in the European Strategy for Particle Physics (ESPP)~\cite{esppu} and the Physics Briefing Book~\cite{pbb}. Thus, in the last ESPP update in 2020, the community was encouraged to define a detector R\&D roadmap, identifying the most important technological developments in the domain of particle detectors required to reach the goals defined in the ESPP. In the period between May 2020 and October 2021, the roadmap was prepared by collecting input from the community, which was eventually collated in a 250-page document and summarised in an 8-page synopsis~\cite{roadmap}.

\section{ECFA Detector Roadmap document}

\begin{figure}[htb] 
\centering
\includegraphics[width=0.95\linewidth]{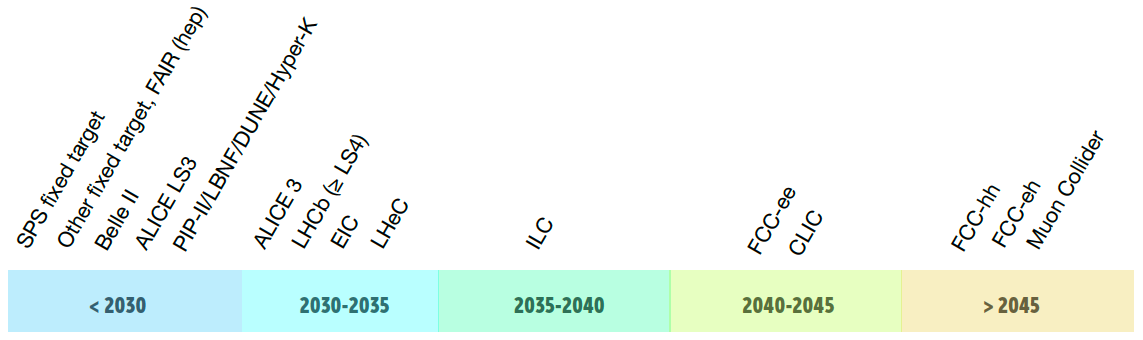}\\\vspace*{2mm}
\includegraphics[width=0.95\linewidth]{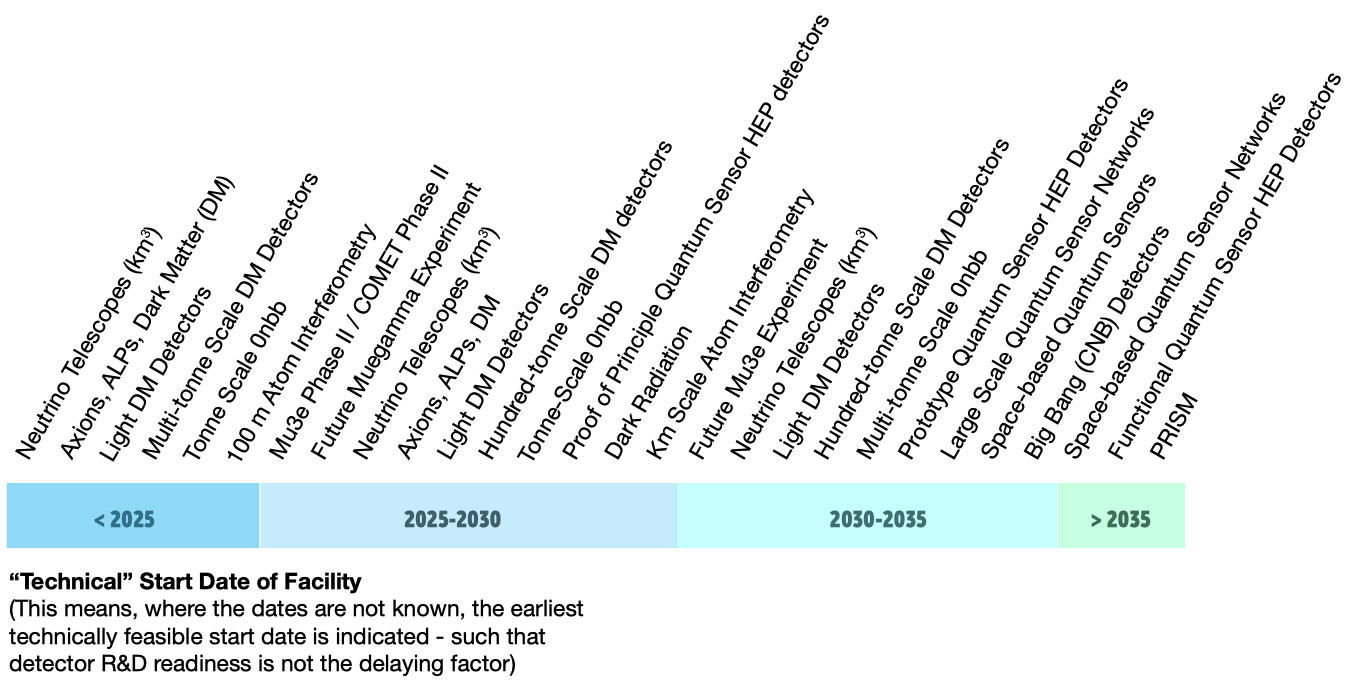}
\caption{Timelines for large HEP experiments (top) and smaller experiments (bottom) as defined in the ECFA Detector Roadmap~\cite{roadmap}.}\label{timelines}
\end{figure}

\begin{figure}[htb] 
\centering
\includegraphics[width=0.95\linewidth]{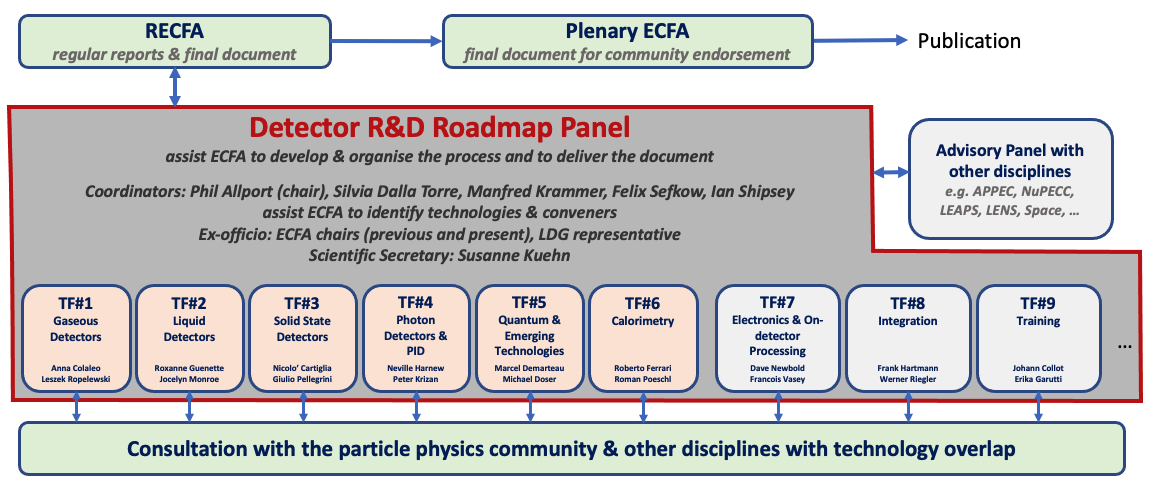}
\caption{Organisational structure and Task Forces (TF) proposed in ECFA Detector Roadmap~\cite{roadmap}.}\label{taskforces}
\end{figure}

Both documents of the ECFA Roadmap contain an overview of detectors for future facilities (such as EIC, ILC, CLIC, FCCee/hh, muon collider), upcoming non-accelerator experiments, major planned upgrades (e.g. ALICE, Belle-II, LHC-b, etc.) and details of the associated timelines (see fig.~\ref{timelines}). 
Six task forces (TF) were defined based on dedicated detector technology domains: gaseous detectors (TF1), liquid detectors for rare event searches and neutrino experiments (TF2), semiconductor detectors (TF3), photon detectors and particle identification (PID; TF4), quantum sensors and emerging technologies (TF5) and calorimetry (TF6). These are complemented by three further task forces, which cover transversal topics related to the detector technologies: electronics and online data processing (TF7), integration (TF8), and training (TF9). 
All TFs and the proposed organization structure from the Roadmap are shown in figure~\ref{taskforces}.

\begin{figure}[htb]
\centering
\includegraphics[width=0.95\linewidth]{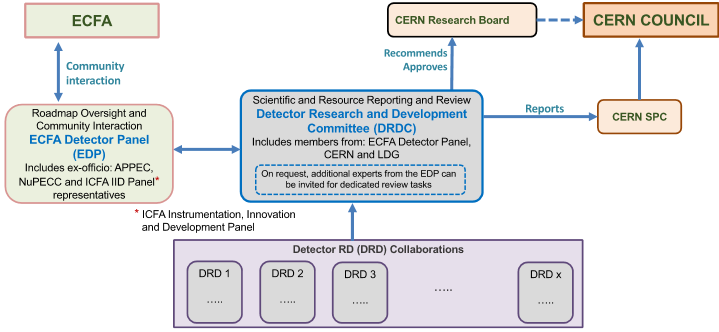}
\caption{Organisational structure for interacting of the DRD collaborations with the DRDC, EDP and CERN Management (the arrows indicate the reporting lines) from \cite{implementation}.}\label{implplan}
\end{figure}

In the autumn of 2022, the CERN Council and Scientific Policy Committee (SPC) endorsed the Detector Roadmap Implementation Plan~\cite{implementation}. This plan suggests that the long-term detector R\&D efforts be organized into larger detector R\&D (DRD) collaborations, each to cover a technology domain identified in the Roadmap, will be hosted at CERN. 
Since TF9 on training does not fit well into this scheme, it was decided to move these efforts into the newly created ECFA Training Panel~\cite{training}.
It is proposed in the Implementation Plan (as outlined in the organigram \ref{implplan}) that a new CERN committee, the Detector Research and Development Committee (DRDC), be established for the purpose of reviewing the new R\&D collaborations, which emerged from the task forces mentioned above. The DRDC is meant to be a new CERN body embedded in the existing CERN committee structure and would ensure rigorous oversight of the process and the DRD collaborations through CERN’s well-known and internationally respected peer-reviewing processes.

The DRDC mandate was developed during the autumn of 2023. It defines the committee's main tasks as receiving and reviewing DRD proposals, suggesting their approval -- when appropriate -- to the CERN Management via the Research Board, and continuously monitoring the progress of each approved collaboration by requesting and reviewing annual status reports. Committee members are appointed by the CERN Director of Research and Computing for a two-year term, which can be renewed up to three times. The DRDC contains experts in different detector technologies, as well as members of the ECFA Detector Panel (EDP), including the Panel’s two co-chairs as ex-offico members, in addition to a representative of the Laboratory Directors Group (LDG). The actual member list can be found on the Committee's webpages~\cite{drdc}.
The DRDC has had a heavy workload from its very beginning since it was only established when the first proposals had already been submitted and were ready to be reviewed, at a time when the mandate and working methods of the Committee and the interfaces with ECFA EDP and other groups had yet to be worked out.

\section{From ECFA Task Forces to DRD Collaborations}
Another community-driven process was started in 2022 by all task forces from the Roadmap process (except TF9 on training, as explained above), aimed to understand better the expertise, needs, and interests of each TF community. To this end, each of the task force convenors organized open meetings, which took place between winter 2022 and spring 2023. Based on the presentations made there and input proposals collected by the TFs, proposal documents for collaborative R\&D in each technology domain were drafted and prepared in spring 2023.

\subsection{First Round of Approvals in December 2023}
Four DRD collaborations submitted their proposals by the deadline end of July 2023. These collaborations are DRD1 (gaseous detectors), DRD2 (liquid detectors), DRD4 (photon detectors and PID) and DRD6 (calorimetry), while DRD3 (semiconductor) submitted its proposal at the beginning of October. 

During nine preparatory DRDC meetings, the five proposals submitted were reviewed internally by the DRDC, which resulted in several suggestions for each DRD proposal writing team. Together with additional direct meetings with the proposal writing teams, this feedback led to updates of the proposals, bringing them towards better consistency with the template provided by EDP and with the expectations of the DRDC. 
One goal of the review was to streamline the contents in terms of terminology. Resource-loaded "work packages" (WPs) will be the main work areas, defining milestones and deliverables and reflecting the strategic funding requests.
Transversal "working groups" (WGs) will cover topics such as infrastructures, software, tools, outreach, and other common activities all WPs need.

At the closed session of the first official DRDC meeting, held on 4 December 2023, each DRD preparation team presented its proposal and answered questions from the Committee. Based on these presentations and the final versions of the proposals, the DRDC drafted recommendations in the meeting minutes, which have been made available via the CERN Document Server (CDS)~\cite{drdccds}.

The proposals and the DRDC recommendations were subsequently presented to the CERN Research Board (RB), consisting of the extended CERN Directorate, which eventually approved the collaborations on gaseous detectors (DRD1), liquid detectors for rare event searches and neutrino experiments (DRD2), photon detectors and particle identification (DRD4), and calorimetry (DRD6) for an initial (and extendable) period of three years. The collaboration on semiconductor detectors (DRD3) was only given conditional approval due to the very late submission of its proposal and several open issues that have not been addressed in time.

Finally, the approved DRD collaborations presented their proposals to the public at the open session of the DRDC meeting on 4 March 2024 and uploaded the proposal documents to the public CDS.

\subsection{Second Round of Approvals in June 2024}

After this initial round of approvals, two further collaborations were approved for three years in a second round in June 2024 by the RB, following reviews and iterations with the DRDC during the months before. These collaborations cover quantum sensors and emerging technologies (DRD5), as well as electronics and on-detector processing (DRD7). Both proto-collaborations submitted a Letter of Intent beforehand about their plans and timeline.

Moreover, full approval for three years was given to DRD3 at the June meetings of the DRDC and RB after the open points had been adequately addressed. 

\subsection{Next Steps after Approval}
The approved collaborations are now shifting from the proposal writing phase to running collaborations. 
They are encouraged to appoint their new management structures quickly, following broad consultation to ensure representation of the whole collaboration. This is especially important for those DRD collaborations that need to ensure the continuation of the ongoing activities of their predecessor collaborations. 
For this, they need to constitute a Collaboration Board with representatives of each collaborating institute. A CB chair election had to be organized by a search committee, following democratic rules and best practices known from other large collaborations. Moreover, a search and election for the collaborations' spokesperson and deputies had to be organized. Some collaborations kept the protagonists from the proposal writing and TF phase, some elected new ones.

Moreover, for each DRD collaboration, all participating institutions must sign a Memorandum of Understanding (MoU). It is based on a template provided by CERN and will be iterated with each collaboration in the next months. It defines the obligations of both CERN as the host laboratory and the collaborating institutions and defines the necessary structures of each collaboration. It defines, in addition to the management structures and Collaboration Board mentioned above, a Resources Board representing the Funding Agencies. It specifies the tasks of Working Groups and Work Packages, regulates intellectual property (IP) topics and describes how to implement a common fund if necessary by each collaboration. It defines the role and obligations of industrial partners and includes CERN's "General Conditions applicable to Experiments" as an underlying framework and integral part.

The main MoU is complemented by a list of annexes, which can be changed by CB decisions and do not require a full signature round. The annexes list the collaborating institutions and their contact persons, funding agencies and their representatives, and the organizational structure of each collaboration. They list collaboration by-laws, which cover rules of the collaborations to regulate themselves, as well as included background IP. 
An essential part are the annexes describing the working groups and the work packages with the resource-loaded milestones and deliverables.

With the initial approval of these collaborations, entries in the CERN Greybook database have been made so that team leaders of participating groups can be appointed and team members can be registered as CERN users.  As a consistent starting point for all meetings and events, respective categories in CERN Indico have been created under the section "Experiments $\rightarrow$ R\&D".
The collaborations are asked to nominate liaisons to other DRD collaborations with topical overlap, as well as to the corresponding CPAD groups in the US~\cite{cpad}.

\section{The new DRD collaborations}

\subsection{DRD1}
Gaseous detectors have a long heritage in HEP following the invention of the multi-wire proportional chamber (MWPC), which shifted particle detection techniques into the electronic era. Further developments, such as drift and time projection chambers (TPC), led to impressive sub-detectors for the LEP experiments. The developments in the last two decades have focused on micro-pattern gas detectors (MPGD) such as GEM and Micromegas, which were further developed and disseminated by the RD51 collaboration from 2008 onwards.

The new DRD1 collaboration is building on these successful developments and increasing the scope to include large-volume tracking systems such as MWPCs, TPCs and RPCs. It comprises 700 participants from 161 institutes in 33 countries, which makes it twice as large as its predecessor, RD51.

\begin{figure}[h]
\centering
\includegraphics[width=0.95\linewidth]{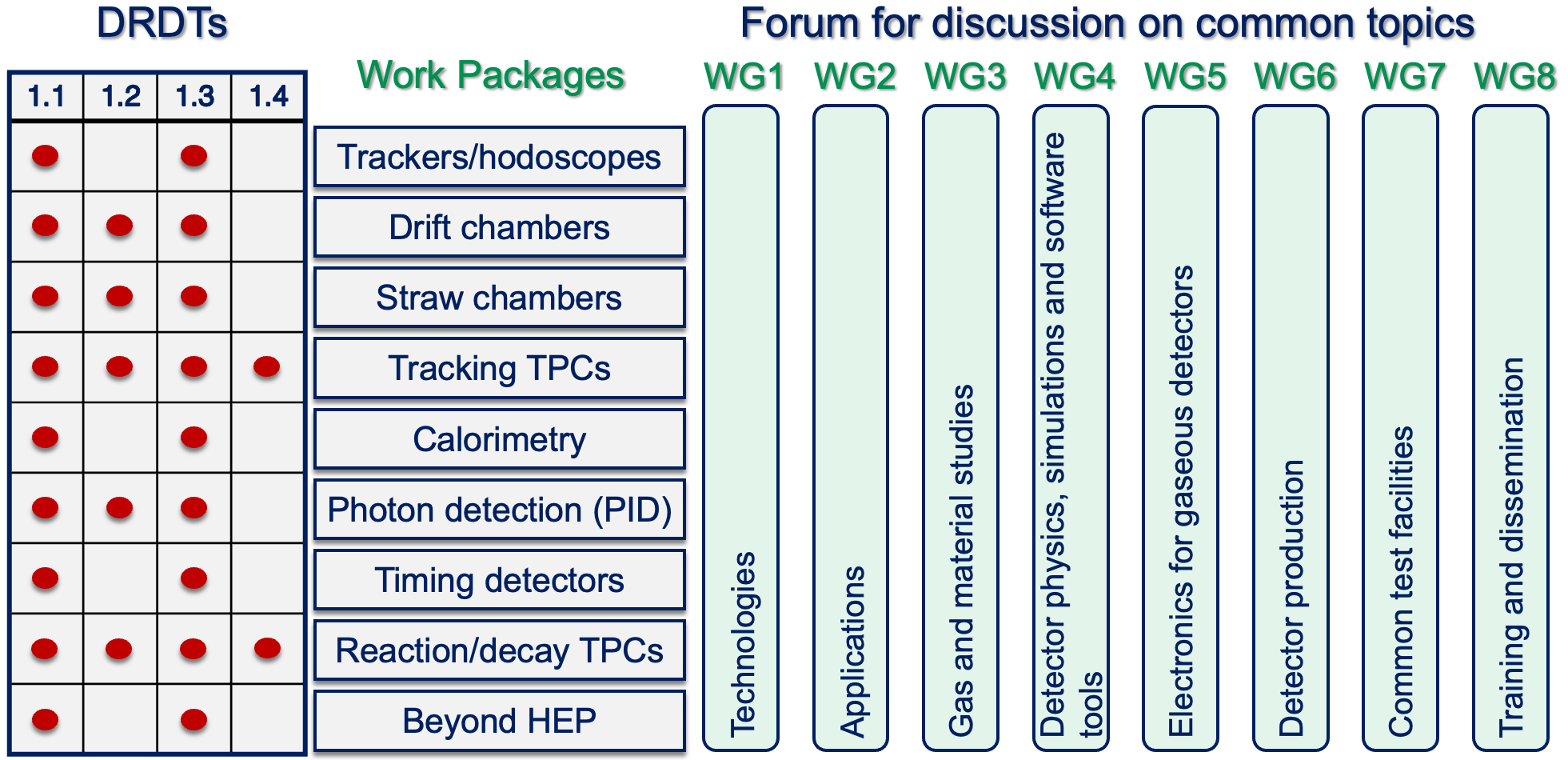}
\caption{Organisational structure of the DRD1 collaboration, showing the Work Packages are transversal Working Groups~\cite{drd1}. The red dots visualize the correlation of WPs with the DRDTs from the ECFA roadmap.}\label{drd1-org}
\end{figure}

DRD1 organized themselves in Work Packages and transversal Working Groups. While the WGs are seen as forums for discussing common topics and serving as the backbone of R\&D, the WPs carry the strategic work program and are linked to the DRDTs from the Roadmap (see fig.~\ref{drd1-org}).
The R\&D covers different technologies reflected in the WPs improving time and spatial resolution for gaseous detectors with long-term stability (DRDT1.1) to achieve tracking in gaseous detectors with dE/dx and dN/dx capabilities in large volumes with very low material budget and different read-out schemes (DRDT1.2). An important sustainability goal is to develop environmentally friendly gaseous detectors for very large areas with high-rate capability in DRDT1.3 and to achieve high sensitivity in both low and high-pressure TPCs (DRDT1.4).

To achieve these goals, the collaboration has 281 FTE manpower and roughly 3MCHF budget available, while an additional 120 FTE personnel and 2.8MCHF additional resources will be needed per year to cover the strategic R\&D (for the initial approved three-year period)

The collaboration held an open meeting on 8 December 2023, where candidates for spokespersons and the CB Chair presented their statements. The collaboration subsequently elected Anna Colaleo (Univ. and INFN Bari) as CB chair and Eraldo Oliveri (CERN) and Maxym Titov (IRFU/CEA) as spokespersons.
The first DRD1 collaboration meeting took place between 29 January and 2nd February of 2024; a second collaboration meeting was held in June.

\subsection{DRD2}
\begin{figure}[htb]
\centering
\includegraphics[width=0.95\linewidth]{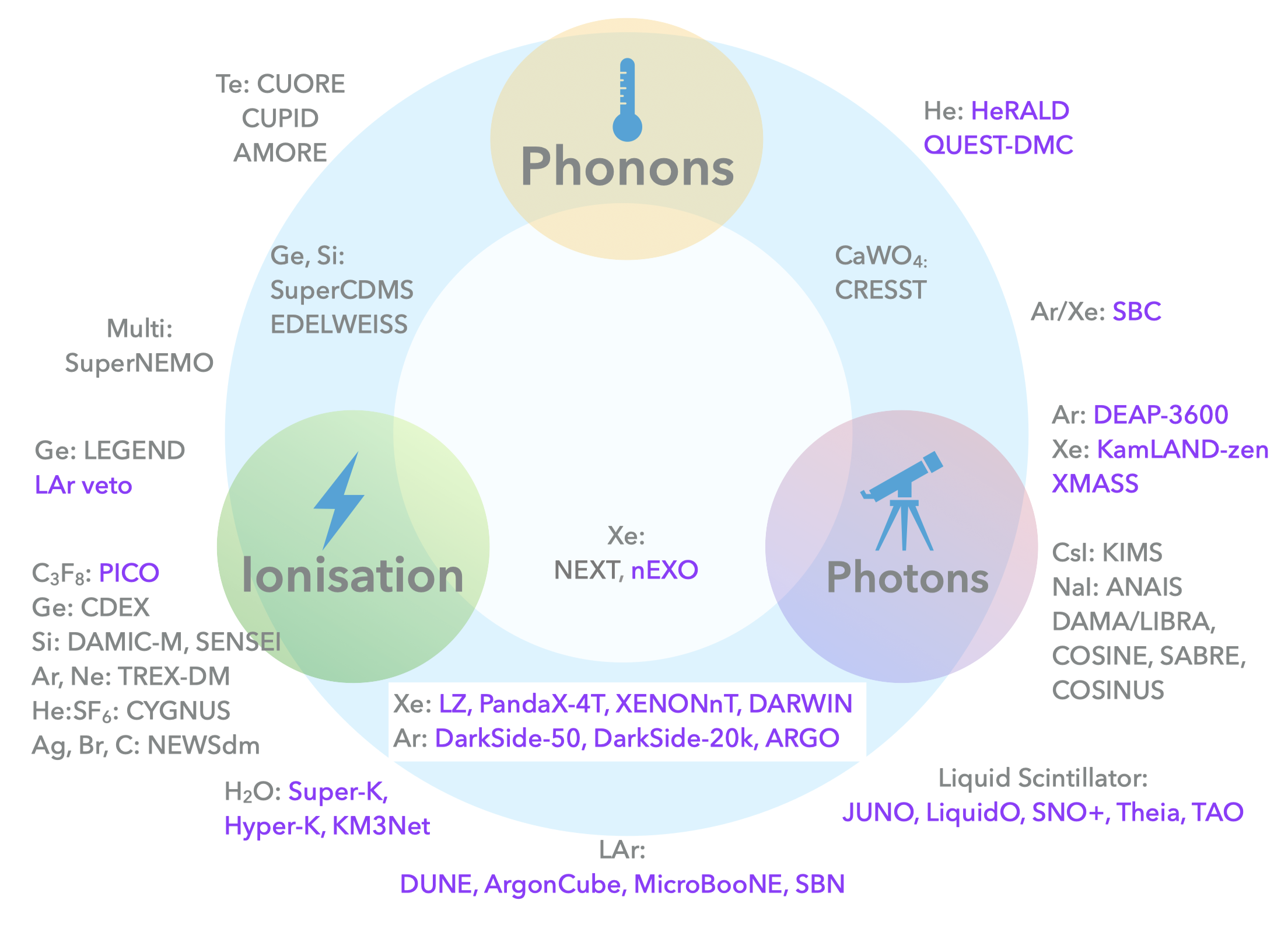}
\caption{Current and near-future experiments addressing the physics drivers for liquid detectors (neutrino physics, neutrino astrophysics, $0\nu \beta \beta$, dark matter), grouped by detection modality, with liquid targets in purple~\cite{roadmap}.}\label{drd2}
\end{figure}

The DRD2 collaboration aims to develop liquid detectors for applications in rare event searches and neutrino physics, in both accelerator and non-accelerator experiments. Several large-scale and many small-scale experiments are running or planned using either noble liquids (e.g. DUNE), water Cherenkov detectors (e.g. Super/Hyper-K) or liquid scintillators with light and ionisation readout, as shown in the sketch in figure~\ref{drd2}. R\&D on target doping and purification is needed for multi-tonne noble liquids, and the radiopurity of detector components and background mitigation also needs to be studied. 
The collaborative effort to build up a research community that has not traditionally worked together so far is to be commended. The DRD2 collaboration is expected to comprise more than 200 participants from 114 institutes in 15 countries, who provide approx. 143 FTE manpower. An additional amount of 300 FTE is claimed to be needed for the work proposed. Budget-wise the collaboration claims to have a budget between 2.6 and 1.5 M€ available per year between 2024 and 2026, while further 8M€ need to be requested to carry out the planned activities. The large difference between available and needed resources for both funds and manpower clearly shows that the collaboration needs to approach their funding agencies to address this.

An inaugural collaboration meeting was held at CERN between the 5th and 7th of February 2024, with 91 contributed talks and 156 participants. The newly formed CB elected W. Bonivento (INFN Cagliari) as chair. Subsequent elections resulted in Giuliana Fiorillo (Univ. Napoli) and Roxanne Guenette (Univ. Manchester) as spokespersons.

\subsection{DRD3}
Semiconductor detectors are a remarkable success for HEP experiments, with sensitive areas having increased by one order of magnitude each decade: 1~$m^2$ (vertex detectors at LEP, e.g. DELPHI as the largest) → 10~$m^2$ (e.g. CDF, where the use of silicon sensors was extended from pure vertexing to tracking) → 200~$m^2$ (CMS Tracker) → 600~$m^2$ (CMS High Granularity Calorimeter currently built for Phase-II Upgrade). 
The DRD3 collaboration~\cite{drd3} aims to continue this success and the work of its predecessors, the RD50 (radiation-hard silicon and other materials), RD42 (diamond) and RD53 (FE-ASICs) collaborations. It is extending the scope even further by an enlarged scope on monolithic CMOS pixel sensors and a new topic on 3D integration, performing R\&D for future lepton colliders, which was not the scope up to now.

\begin{figure}[htb]
\centering
\includegraphics[width=0.95\linewidth]{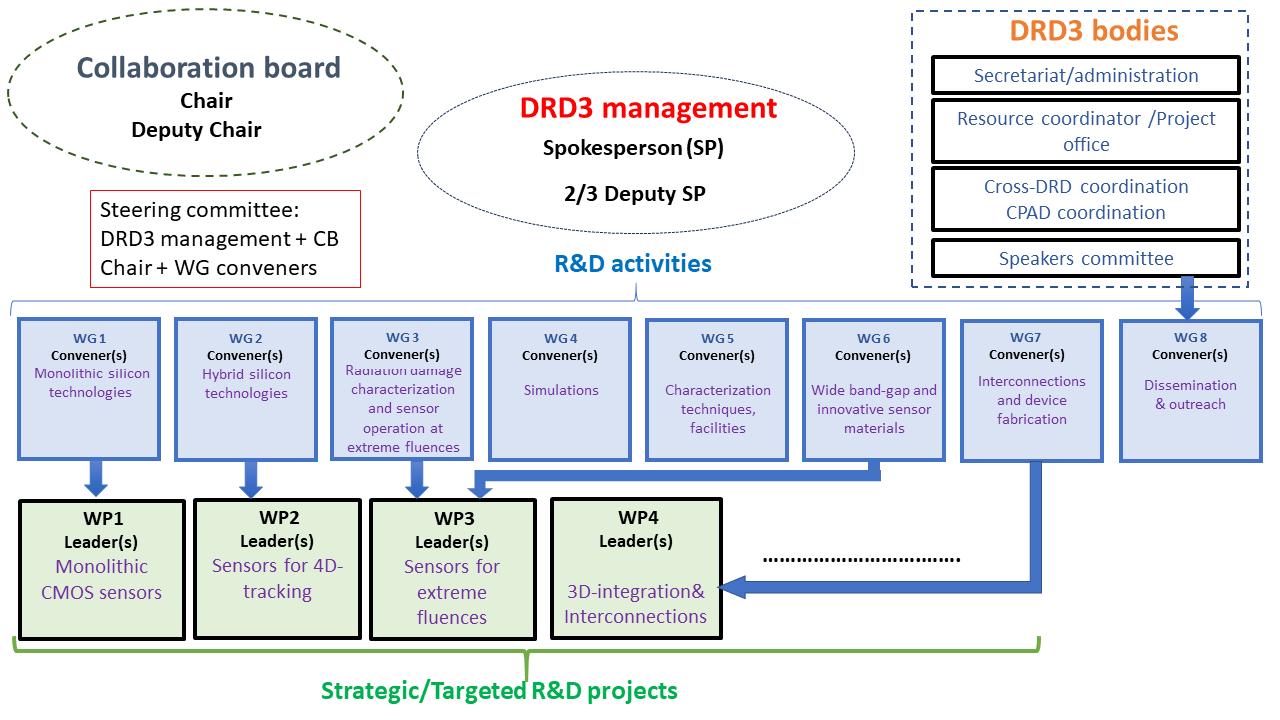}
\caption{Organizational chart of the DRD3 collaboration from its proposal document~\cite{drd3}.}\label{drd3}
\end{figure}

As the proponents state that DRD3 serves a dual purpose of promoting strategic developments as described in the Roadmap~\cite{roadmap}, while promoting blue-sky R\&D in solid-state detectors, they strongly emphasize Working Groups. Their organization structure (shown in fig.~\ref{drd3}) focuses on eight working groups. Three are a full replica of the corresponding Work Package, namely WG1/WP1 on monolithic CMOS sensors, WG2/WP2 on hybrid sensors for 4D tracking and WG7/WP4 on 3D integration and interconnection techniques. Only WP3 on sensors for extreme fluences is fed by two Working Groups, namely WG3 on radiation damage and WG6 on wide-bandgap material.

There are more than 900 interested people from 132 institutes in 28 countries, making it twice as large as RD50, with 70~\% from Europe. They generate 327 FTE of manpower, while another 170 FTE is needed to fulfill the proposed work. The available budget is 5MCHF per year, with an additional need of 7.9MCHF to be requested.

After elections, Giulio Pellegrini (IFIC-CNM Barcelona) acts as CB chair, with Roberta Arcidiacono (Univ. Piemonte \& INFN Torino)  as his deputy. The management of the collaboration is led by Gregor Kamberger (JSI Ljubljana) as spokesperson, accompanied by the deputies Michael Moll (CERN), Ingrid-Maria Gregor (DESY), and Sally Seidel (Univ. of New Mexico), who also acts as one of the two chairperson of the corresponding US RDC3 collaboration on solid state tracking (together with Anthony Affolder/UCSC). Convenors for all WG have been nominated. Some of them also act as WP leaders of the corresponding WPs.

The first full collaboration meeting was held between the 17 and 21 of June 2024, and the second one is planned for the beginning of December. In between, the WG convenors have been nominated and started calling for working group meetings.

\subsection{DRD4}
The DRD4 collaboration aims to develop photon detectors and particle identification devices. It comprises participants from 74 institutes in 19 countries, including many small groups with no prior experience in large collaborations. Nevertheless, the community-building effort, bringing these groups together, has been seen as very positive and concluded in a strong proposal~\cite{drd4}, coordinated by Christian Joram (CERN) and Peter Kri\v{z}an (JSI Ljublijana). 

Scientifically, the work focuses on the development of photon detectors like MCP-PMTs, SiPMs, vacuum, and gaseous photon detectors and their application in ring imaging Cherenkov detectors (RICH), time-of-flight (ToF) and transition radiation detectors (TRD). 
The work is organized in five WPs, which follow the DRDT themes in the Roadmap~\cite{roadmap}: WP1 on solid-state and WP2 on vacuum-based photodetectors. WP3 is about RICH detectors and WP4 on Time-of-Flight based systems. WP5 covers advancements in scintillating fiber (SciFi) and TRDs. 
The WGs address the thematic fields: all photodetector technologies (WG1), Particle ID using RICH, TOP, TOF and other methods (WG2), Technologies like radiators, optical elements, readout and cooling (WG3), simulation and reconstruction software (WG4). WG5 resembles WP5 on SciFi and TDRs, and WG6 covers novel ideas and blue-sky research.

59 groups participate in one or several work packages and foresee investing resources. In the first three years, an annual labor force of around 100 FTEs will be available, with an estimate of additional 50 FTEs per year required to fulfil the tasks as planned.
The available financial resources amount to approximately 1.5 M€ per year, with an additional 2 M€ per year considered necessary for the efficient and timely implementation of the work program.

The collaboration held a constitution meeting on 23 and 24 January 2024 to elect Massimiliano Fiorini (Ferrara) as spokesperson and Guy Wilkinson (Oxford) as CB chair, as well as most WG and WP convenors. The first full collaboration meeting was held between 17 and 20 of June this year.

\subsection{DRD5}
The unprecedented sensitivity and precision of quantum systems enable the investigation of fundamental questions in particle physics. Quantum technologies are also rapidly emerging in various domains nowadays. 
The DRD5 collaboration follows these advancements and aims to identify key technologies and topics that will benefit our community the most. Three domains are being followed: a) exploring fundamental physics questions, such as demonstrating fundamental symmetries through particle, atomic, or molecular EDM, spectroscopy, and the search for dark matter and new couplings, b) enhancing quantum measurements by addressing foundational questions and testing the violation of fundamental symmetries and interactions, and c) developing applied technologies for novel types of detectors using new materials and extreme sensitivity through phase transitions.

Since there has been no large collaboration working on these topics before, another important goal of DRD5 is to identify groups willing to participate in a global collaborative effort and establish trust and mutual interest to identify benefits across these groups. For this, community building is a very important aspect. Thus, no membership fee or common fund is foreseen, and joining the collaboration will be lightweight via a simple request to the collaboration board. 
The proposal~\cite{drd5}, developed in the last months, was submitted to the DRDC at the end of February 2024 and sent to interested institutions in parallel. From all responses, 94 groups with 338 individuals have expressed interest in signing the proposal for its approval, while the actual FTE number is assumed to be only a fraction. 

\begin{figure}[htb]
\centering
\includegraphics[width=0.95\linewidth]{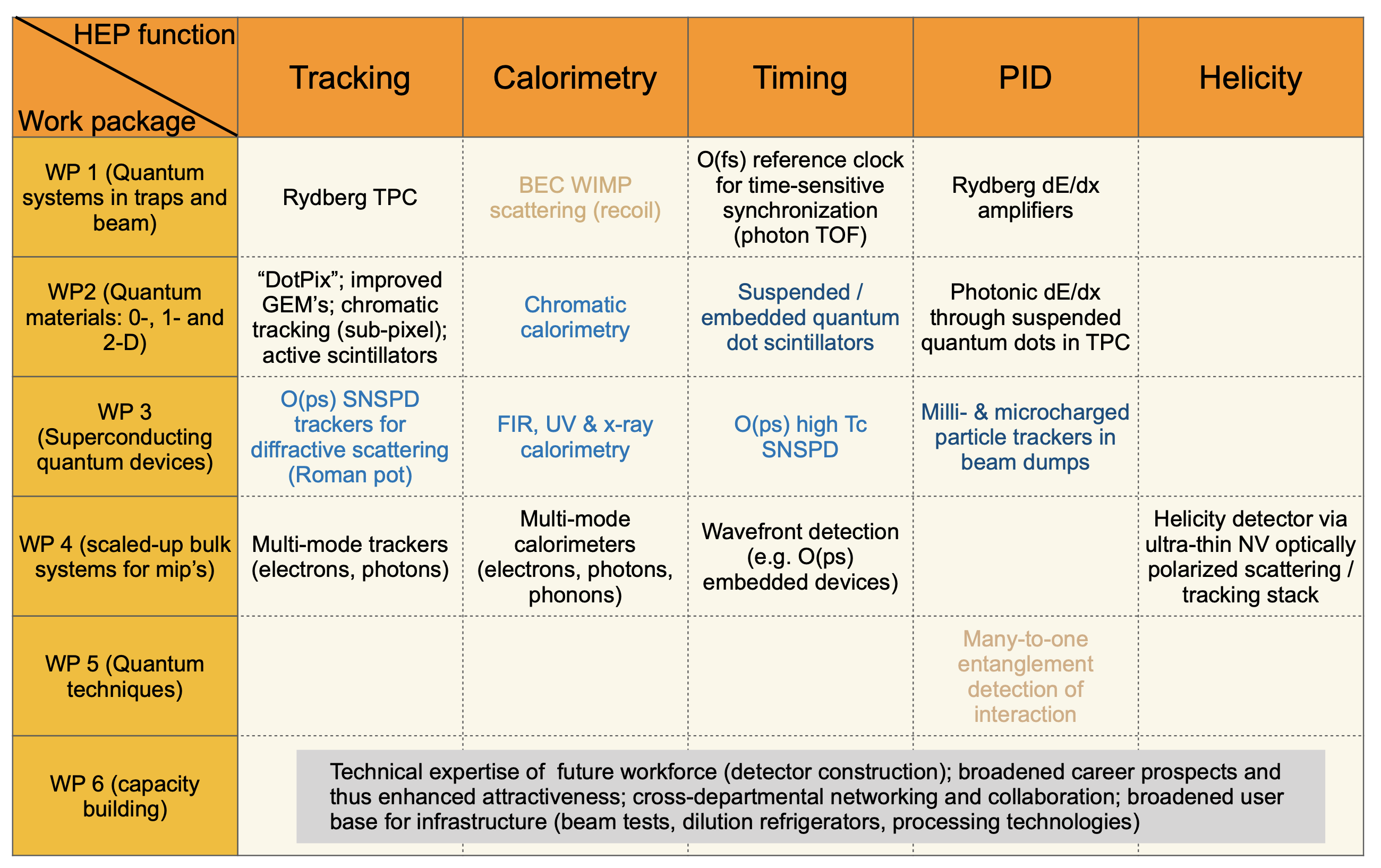}
\caption{Application of quantum sensor technologies investigated within the work packages of DRD5 for different HEP detectors~\cite{drd5}.}\label{drd5}
\end{figure}

Based on the Detector R\&D Themes (DRDTs) and topics covered in the ECFA Detector Roadmap, the DRD5 collaboration has identified six work packages: Work Package (WP) 1 covers exotic systems in traps and beams, atom interferometry, as well as networks, signals, and clock distribution. WP2 is about building blocks for complex nanoscale low-dimensional "quantum materials". It includes application-specific research, such as investigating quantum dots and monolayers with extended functionality, and its inclusion in simulation packages, such as Geant4. WP3 deals with cryogenic materials, devices, electronics, and systems, and its aim, among others, is to investigate the role of high-$T_c$ superconductors in HEP instrumentation. WP4 covers large spin-polarized ensembles, molecules with radioisotopes for EDM searches, hybrid devices like quantum wells in semiconductors, or quantum cascade lasers coupled to silicon sensors, as well as opto-mechanical sensors for DM, neutrinos, and gravitational wave searches by using low-loss solids, levitated particles or sensors based on superfluidity. WP5 deals with opto-mechanical resonators and atom interferometers for GW detection and beyond, as well as investigation of the use of quantum entanglement in HEP. In WP6, education and exchange platforms, as well as infrastructures, are being covered.
Figure~\ref{drd5} shows which potential application in HEP the different work packages could have.

The milestones and deliverables in each WP are mostly reports and documents and fewer physical objects, as this proposal covers much lower Technology Readiness Levels (TRLs) than other DRD collaborations. Thus, it is accepted for the time being that neither resource planning was done for the deliverables nor institute responsibilities were defined yet.  In its meeting in June, the DRDC suggested selecting the more mature projects and starting such resource planning, together with establishing contacts with the other DRDs on possible areas of overlap.

\subsection{DRD6}
R\&D in calorimetry has a particularly long lead-time due to many readout technologies developed elsewhere, utilizing gas-, scintillator-, and silicon-based detectors. Another factor are the large and challenging prototype set-ups, even in the early stages. Moreover, future colliders have high expectations on energy and time resolution (DRDT6.1), as well as granularity allowing particle flow methods (DRDT6.2), rate \& pileup compatibility and radiation tolerance (DRDT6.3).

Thus, the DRD6 collaboration aims to bring a diverse set of calorimeter technologies to a level of maturity such that they can be considered for a technology selection of future experiments. The work is organized into four Work Packages: WP1 covers electronics and service integration aspects of sandwich calorimeters and is explained in more detail in fig.~\cite{drd6}. WP2 deals with liquified Noble gas calorimeters, while developing materials and photodetector readout systems for scintillators is examined in WP3. In addition, there is a dedicated WP4 on front-end ASIC developments, as expensive wafer productions are planned to be covered by strategic funds.
\begin{figure}[htb]
\centering
\includegraphics[width=0.95\linewidth]{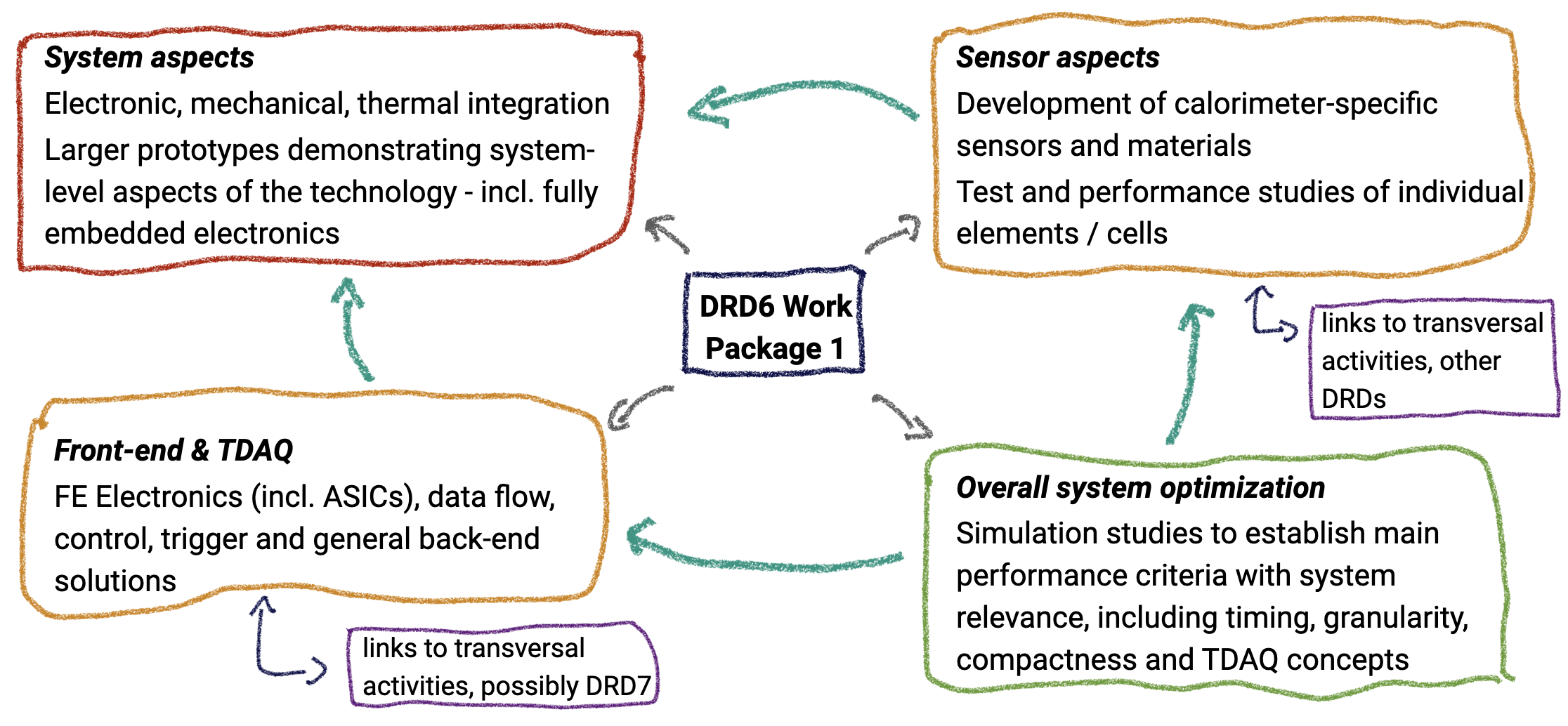}
\caption{Synthetic overview on the main research directions common to all projects in WP1 of DRD6 on calorimetry~\cite{drd6}.}\label{drd6}
\end{figure}
Working Groups complement the WPs on photodetectors (WG1) with synergies to DRD4. In WG2, on testbeam infrastructure, a dedicated beamline at the SPS is planned to be built and operated by the collaboration in the CERN North Area. Simulations, algorithms, and software tools will be further developed in WG3. The WG4 on industrial connection and Knowledge Transfer (KT) aims to foster collaboration with companies producing materials, photodetectors and electronics used in calorimetry. Finally, a WG5 was foreseen on mechanics and integration if the future DRD8 collaboration does not cover this topic.

The DRD6 collaboration has connections to various groups and collaborations, including CALICE, CrystalClear (CERN RD18), FCal, GranuLAr, CalVision, and EU projects such as AIDAinnova, EuroLabs, the CERN EP-R\&D program, and several proto-collaborations like ILC, CLIC, or FCC. Some of these collaborations continue to exist alongside DRD6, even if they are focusing on calorimetry: CALICE will continue in "hibernation" to keep the speakers bureau and other structures. CrystalClear has a broader scope beyond high-energy physics, with developments used in medical applications such as Positron Emission Tomography (PET).

The DRD6 collaboration comprises participants from 132 institutes in 29 countries. The available manpower is around 183 FTE per year, with an additional 100 FTE needed to fulfill the research goals. The anticipated annual budget is around 3.2M€ of existing funds, with 1.4 to 2.4M€ per year of additional needed funds between 2024 and 2026.

DRD6 held a proto-collaboration meeting mid-January 2024, where decisions on collaboration bodies have been made. A collaboration kick-off and first CB meeting was organized on 9 -- 11 April 2024, with 133 registered participants. Elections resulted in Roberto Ferrari (INFN Pavia) as CB chair and Roman Pöschl (IN2P3-IJCLAB) as the spokesperson. Both were already involved in the proposal writing team in leading roles. 

\subsection{DRD7}
High-performance electronic systems are an essential aspect of all HEP experiments. The complexity and costs of the necessary developments are high and will continue to increase.
The objective of the DRD7 collaboration~\cite{drd7} is to carry out strategic R\&D in electronics, fulfilling the DRDTs from the ECFA Roadmap. TF7 (now DRD7) was listed there, together with integration/mechanics (TF8) and education/training (TF9), as an orthogonal task force, since these topics are essential for all DRD collaborations. However, the DRD7 collaboration has been set up to perform R\&D on electronics, covering low TRLs 1-5 and targeting disruptive, transformative, and mid-to-long-term goals. It does not act as a "service provider" for other DRDs, where higher TRL components are needed on shorter time scales.
However, DRD7 will also provide access to expertise, tools, and industry vendors, supporting the entire DRD program and facilitating future standards and approaches. It will support both specific technical goals in the area of electronics and the general strategic recommendations of the Roadmap. Thus, interfaces to other DRDs are being defined, and liaison persons have been nominated.

The proto-DRD7 collaboration had an open workshop in March 2023, where WP convenors were nominated, and proto-projects were defined, which were reviewed and refined in a second workshop held in September 2023, when also a Letter of Intent was submitted to the DRDC.
The full proposal of the DRD7 collaboration was submitted to the DRDC in February 2024. After review and iterating with the proponents, it was presented at the open session of the DRDC meeting and approved by the CERN Research Board in June this year. A first full collaboration meeting will be held on 9-10 Sept 2024 at CERN. In parallel, a collaboration board, comprising representatives of each of the 68 member institutes from 19 countries, is forming with a chair election. The first proto-CB meeting was held on 26 June. 
DRD7 has chosen a different management way than the other collaborations, comprising a steering committee rather than a single spokesperson as a central executive body. The steering committee members, which will be replaced on a rolling basis, currently comprise Francois Vasey (CERN), Frank Simon (KIT), Jerome Baudot (IPHC Strasbourg), Marcus French (RAL), Ruud Kluit (NIKHEF), and Angelo Rivetti (INFN Torino). The first two act as co-chairs and represent the collaboration to the outside. 

Based on the five DRDTs from the Roadmap, DRD7 has defined six genuine research WPs and one transversal one on tools and technologies (WP7) based on proposed projects in a bottom-up approach. However, certain DRDT topics have not been covered so far, in particular intelligent power management, data reduction techniques based on AI, novel on-chip architectures, as well as reliability and fault tolerance.
WP1 covers data density and power efficiency topics, targeting e.g. silicon photonics transceivers, power systems and wireless transmissions. WP2 deals with radiation-tolerant RISC-V System-on-Chip ASICs to increase the intelligence in the detector. WP3 aims to develop high-performance TDC and ADC blocks for ASICs with ultra-low power and high-precision timing distribution. WP4 is on extreme environments, covering the development of CMOS PDKs at cryogenic temperatures and research on radiation-resistant CMOS nodes, as well as cooling (which partly overlaps with topics of DRD8 in the future). WP5 works on DAQ platforms and simplified backend systems and investigates commercial off-the-shelf electronics. WP6 deals with imaging ASICs for MAPS sensors and technologies, providing common access to selected CMOS nodes like TPSCo 65 nm, Tower 180 nm, and LFoundry 110 nm. WP7, as mentioned above, proposes a hub-based structure to support ASIC developments of the HEP community in general and the other DRD collaborations in particular, establishing and maintaining access to process nodes at large foundries and EDA software tools and facilitating collaborative work like for IP-block sharing. This WP heavily relies on collaboration and coordination with experienced regional centers (Hubs) and on structures set up in the European context, such as Europractice IC and Software services. Internationalizing these efforts will be challenging due to licensing and legal topics.

The committed manpower in all WPs is 110 FTEs on average for the next three years. Around 67 FTEs are needed annually in addition to achieve the planned work program. This might sound less than for the other DRD collaborations. However, DRD7 has categorized its membership into contributors who provide resources to dedicated tasks (224 people) and observers who only follow the activities in the WPs without commitments (currently 178 names). In the other DRD collaborations, this distinction has not been made so strict and must be considered when compared.

\subsection{DRD8}
As the initial TF8 convenors did not continue as proposal preparation team, new convenors had to be searched for, which were found around the group organizing the "Forum on Tracker Mechanics" workshop series. A survey was conducted, and it revealed that the community is interested in moving forward. Subsequently, an open meeting was held in December 2023 at CERN, and the DRDC received a {\it Letter of Intent} aiming to submit a full DRD8 proposal by the end of this year, which is coordinated by a steering group comprising A. Jung (Purdue), A. Mussgiller (DESY), C. Gargiulo, P. Petagna, B. Schmidt (all CERN) and G. Viehhauser (Oxford). However, the {\it LoI} does not cover all DRDTs of TF8, as they are quite diverse and focus on vertex detector mechanics and cooling only, with interest from currently 22 institutes in seven countries, while magnet development, as well as beam, radiation, and environment monitoring not covered.

\section{Summary and Outlook}
The CERN Research Board approved the first DRD collaborations in December 2023 after a thorough review and iterations with the DRDC, the newly established Detector R\&D Committee.
These collaborations cover gaseous detectors (DRD1), liquid detectors for rare event searches and neutrino experiments (DRD2), photon detectors and particle identification (DRD4), and calorimetry (DRD6). 
After this initial round of approvals, two further collaborations were approved in a second round in June 2024, following similar procedures. These collaborations cover quantum sensors and emerging technologies (DRD5), as well as electronics and on-detector processing (DRD7). Moreover, full approval for three years was given to DRD3 on semiconductor detectors after some open points in an approval request at the first round had been adequately addressed. 

Apart from Task Force 9, which became the ECFA Training panel, all other topics listed in the ECFA Roadmap are now actively followed in DRD collaborations except TF8 (Integration). However, there is the intention to submit a proposal for a DRD8 collaboration on mechanics and colling before the last DRDC/RB meetings in November/December 2024.

The collaborations are now starting their operational phase. In parallel, each collaboration is working on a Memorandum of Understanding. Based on a CERN template, this text is the basis of the collaboration and is aimed at being signed by every participating institution. Apart from organizational matters, it also defines the work packages and resource commitments for deliverables described in an MoU annex.

The DRD collaborations are approved for an initial period of three years. The DRDC will call for annual reports in the DRDC meetings, with 20-minute presentations in the open session on scientific progress and a short discussion on organizational matters (MoU,...) in the closed session. 
The next DRDC meeting will happen in November this year. Its open session will have presentations of all DRD collaborations approved in the first round, presenting the progress from their first year (DRD1, DRD2, DRD4 and DRD6). A similar session is planned for spring 2025 with presentations of the collaborations (fully) approved in the second round, i.e. DRD3, 5 and 7.
Longer presentations, as well as written status and prolongation reports, will be requested before the end of the three-year approval period.

\end{document}